\documentclass[12pt]{iopart}
\usepackage{iopams}
\usepackage{latexsym}
\usepackage{graphicx}
\usepackage{wrapfig}
\usepackage{color}
\usepackage{cite}
\renewcommand{\eref}[1]{Eq.~(\ref{#1})}
\renewcommand{\fref}[1]{Fig.~\ref{#1}}
\def\dfrac#1#2{{\displaystyle {#1 \over #2}}}
\def\pte  {p_\mathrm{te}}
\def\uth  {u_\mathrm{th}}
\def\ute  {u_\mathrm{te}}

\def\vti  {v_\mathrm{ti}}

\begin{document}

%==========================================================================

\title[On stability of collisional coupling]
{On stability of collisional coupling between relativistic electrons 
and ions in hot plasmas}

\author{I.~Marushchenko$^1$, N.~A.~Azarenkov$^1$, N.~B.~Marushchenko$^2$}

\address{$^1$ V.~N.~Karazin Kharkiv National University, Svobody Sq.~6, 61022, 
Kharkiv, Ukraine}

\address{$^2$ Max Planck Institute for Plasma Physics, EURATOM Association,
         Wendelsteinstr.~1, 17491 Greifswald, Germany}

%\ead{nikolai.marushchenko@ipp.mpg.de}

\begin{abstract}
The collisional coupling of relativistic electrons and non-relativistic 
ions in hot plasmas has been analysed.
It is found that relativistic effects produce a new feature:
while the condition $T_e<3T_i$ guarantees a stable collisional 
coupling between electrons and ions in low-temperature plasmas, relativistic 
effects shift the upper $T_e/T_i$ boundary of stability to higher values.
Moreover, for sufficiently high temperatures, $T_{e,i}>75$~keV, collisional 
decoupling between relativistic electrons and ions becomes impossible.
\end{abstract}

%\keywords{hot plasmas, relativistic approach, collisional energy exchange}

\maketitle

%==========================================================================
% paperbody:
%==========================================================================
\newpage
%=======================================================================
\section{Introduction}
\label{sec:intro}
%=======================================================================

Relativistic plasma effects were recognized as important in astrophysics 
long ago and the necessary formalism for description of relativistic 
plasmas has been developed 
\cite{BeliaevBudker:DAN1956mix,deGroot:RelativKineticTheory_1980}.
Furthermore, the production and heating of plasmas by high-power laser 
pulses has been intensively studied 
\cite{Dunne:LaserFusionFacilityEurope2006}, 
and for the interpretation of experimental results a relativistic treatment 
was recognized to be necessary
\cite{HondaMima:JPSJ1998,RobicheRax:PRE2004}.
The progress in fusion, where high temperatures are mandatory 
\cite{Ward:PPCF2010,Wagner:PPCF2010,Stott:PPCF2005},
also makes the kinetics in hot plasmas a subject of importance, 
where relativistic effects sometimes need to be taken into account. 
So far, only a few problems related to relativistic effects in hot fusion 
plasmas have been considered: in particular, solving of the relativistic 
Spitzer problem for calculation of conductivity \cite{BraamsKarney:PoFB1989} 
and of current drive efficiency (see Ref.~\cite{Marushchenko:PoP2011} 
and the references therein).
Apart from this, some aspects of the transport theory in relativistic 
plasmas were considered in Ref.~\cite{MettensBalescu:PoF1990}, 
where the covariant formulation with 4-vectors was applied.

There is a widespread opinion that relativistic effects in laboratory 
devices are important only with respect to the populations of highly energetic 
electrons (see, for example, Ref.~\cite{Ward:PPCF2010}). 
However, relativistic effects can appear also due to the macroscopic 
features of the relativistic thermodynamic equilibrium given by 
the J\"uttner distribution function \cite{deGroot:RelativKineticTheory_1980}, 
also known as the relativistic Maxwellian \cite{BraamsKarney:PoFB1989}.
In particular, contrary to the non-relativistic Maxwellian, the shape of 
the J\"uttner-Maxwellian distribution function depends on the temperature, 
and is Gaussian only in the non-relativistic limit.

As is well known from experiments in toroidal devices 
\cite{BraunEmmert_NF1984,WagnerStroth_Transport-expview:PPCF1993,Zhang_HT7:JPP2010},
electrons and ions exchange energy through collisions, but if the electrons 
are much hotter than the ions the two species ``decouple".
The collisional decoupling thus has not only the heating-power threshold, but 
also a threshold with respect to the ratio of electron and ion temperatures. 
Actually, the value of this temperature threshold depends on the transport 
phenomena (for example, the ``electron root'' in stellarators 
\cite{Maassberg_eroot:PoP2000,Rome_eroot:PPCF2006} establishes a large 
positive radial electric field together with large $T_e/T_i$ ratio, i.e. 
with $e-i$ collisional decoupling).
Apart from this, the radiative losses (bremsstrahlung and cyclotron radiation) 
\cite{Bekefi:RadProc_1966} become important for high temperatures in the energy 
balance of electrons. 
However, only the collisional channel of energy transfer from electrons 
to ions defines the minimal value of the temperature threshold of the decoupling, 
while the balance of heating and losses is responsible only for establishment of 
the steady state.

In this paper, the influence of relativistic effects on the temperature 
threshold with respect to collisional decoupling is studied.
This investigation is applicable to the case when the heating of electrons 
by any external source (for example, high-power laser-pulses in implosive 
plasmas or radio-frequency wave-beams in toroidal fusion plasmas) is balanced 
predominantly by the collisional energy-exchange with the ions (which frequently 
is the most desirable case for experiments without direct heating of ions).
The opposite case, when the collisional transfer of power is of minor importance 
and heating of electrons is balanced by transport and/or radiation losses, 
is usually undesirable and is not considered here. 

%=======================================================================
\section{Energy balance in relativistic plasmas}
\label{sec:balance}
%=======================================================================

Let us assume that electrons and ions have their own Maxwellians with 
the temperatures defined by the energy balance (this is true if the rate 
of thermalization within each of the plasma components is sufficiently 
high in comparison with the external heating). 
The Maxwellian for the ions with density $n_i$ and temperature $T_i$ is 
taken as the classical one,
\begin{equation}\label{Maxwell}
f_{iM}=\dfrac{n_i}{\pi^{3/2}\vti^3}e^{-v^2/\vti^2},
\end{equation}
where $\vti=\sqrt{2T_i/m_i}$ is the ion thermal velocity, while the electrons 
with density $n_e$ and temperature $T_e$ are considered as relativistic 
with J\"uttner-Maxwellian distribution function
\cite{deGroot:RelativKineticTheory_1980,BraamsKarney:PoFB1989},
which it is convenient to represent as
\begin{equation}\label{Juttner}
f_{eJM}=\dfrac{n_e}{\pi^{3/2}\ute^3}C(\mu_r)\,e^{-\mu_r(\gamma-1)},
\end{equation}
where $\ute=\pte/m_{e0}$. Here, $\pte=\sqrt{2m_{e0}T_e}$ is the thermal 
momentum, $m_{e0}$ is the rest-mass, $\mu_r=m_{e0}c^2/T_e$ and 
$\gamma=\sqrt{1+u^2/c^2}$ is the Lorentz factor with momentum per 
unit mass $u=v\gamma$.
Since $f_{eJM}$ is normalized by a density, $n_e=\int f_{eJM}d^3u$,
\begin{equation}\label{CM}
C(\mu_r)=\sqrt{\dfrac{\pi}{2\mu_r}}\dfrac{e^{-\mu_r}}{K_2(\mu_r)}
\simeq 1-\dfrac{15}{8\mu_r}+...\;\;\;(\mu_r\gg 1),
\end{equation}
where $K_n(x)$ is the modified Bessel function of the second kind of 
the $n$-th order.
Here, it is appropriate to recall that contrary to the classical Maxwellian,
the shape of which is independent of the temperature, the relative ``weight'' 
of electrons with $u/\ute\gg 1$ increases at high $T_e$, which shifts the 
``centre of mass'' of the J\"uttner-Maxwellian from the bulk to the tail.

The energy-balance equation in relativistic plasmas can be presented 
in the same form as in non-relativistic ones.
By weighting the relativistic kinetic equation for electrons 
\cite{BraamsKarney:PoFB1989} with the energy $m_{e0}c^2(\gamma-1)$ 
and integrating over momentum, one can obtain the energy balance equation 
(for simplicity, all terms related to inhomogeneity are omitted),
\begin{equation}\label{e-balance}
\dfrac{\partial W_e}{\partial t}=
P_{ei}+P_\mathrm{ext}-P_\mathrm{loss},
\end{equation}
where $W_e=\int m_{e0}c^2(\gamma-1)f_{eJM}d^3u$ is the energy enclosed in 
the relativistic J\"uttner-Maxwellian, 
$P_{ei}$ is the rate of energy exchange between relativistic electrons and 
classical ions, $P_\mathrm{ext}$ is the power of external heating and 
$P_\mathrm{loss}$ is the total loss including transport and radiative losses.
A similar equation can be written also for the ions with $P_{ie}=-P_{ei}$.

Note, that the balance of $P_\mathrm{ext}$ and $P_\mathrm{loss}$ is important 
only for establishing the plasma temperature, and the temperature dependence 
of $P_{ei}$ is the only factor responsible for appearance of the phenomenon 
known as ``collisional decoupling'', which is considered below.
It can be mentioned here also that the losses due to bremsstrahlung 
\cite{Bekefi:RadProc_1966}, $P_{Bs}\propto\sqrt{T_e}$, 
which are increasing with $T_e$, do not produce any effect on the collisional 
energy transfer to the ions and formally it is assumed that $P_{Bs}$ is 
included in $P_\mathrm{loss}$.

%=======================================================================
\section{Collisional energy exchange in relativistic plasmas}
\label{sec:exchange}
%=======================================================================

The rate of energy exchange between relativistic electrons and ions 
with Maxwellian distribution functions, 
$P_{ei}=m_{e0}c^2\int (\gamma-1)C_{ei}[f_{eJM},f_{iM}]d^3u$,
can be written for the relativistic collision operator 
\cite{BraamsKarney:PoFB1989} as follows:
\begin{equation}\label{Pei2}
P_{ei}=4\pi m_{e0}\dfrac{T_e-T_i}{T_e}\int_0^\infty\dfrac{u^3}{\gamma}
F_u^{e/i}(u)f_{eJM}(u)du,
\end{equation}
where $F_u^{e/i}(u)$ is the Coulomb drag of the relativistic electrons 
with ions. 
Since $m_i\gg m_{e0}$, the main contribution in the integral in \eref{Pei2} 
comes from the range $u\gg v_{ti}$ where $F_u^{e/i}(u)$ can be approximated
\cite{BraamsKarney:PoFB1989} as
\begin{equation}\label{Fuei}
F_u^{e/i}(u)\simeq -\nu_{e0}\uth^3\dfrac{n_i Z_i^2}{n_e}\,\dfrac{m_e}{m_i}
\,\dfrac{\gamma^2}{u^2}
\end{equation}
with $\nu_{e0}=4\pi n_e e^4\ln\Lambda/(m_{e0}^2\uth^3)$. 
The final expression for the rate of $e-i$ energy exchange is 
\cite{BeliaevBudker1956_misprnt}
\begin{equation}\label{Pei3}
P_{ei}= P_{ei}^{(cl)}\,                
C(\mu_r)\left(1+\dfrac{2}{\mu_r}+\dfrac{2}{\mu_r^2}\right),
\end{equation}
where $C(\mu_r)$ is defined by \eref{CM} and $P_{ei}^{(cl)}$ is 
the classical expression for the rate of collisional energy exchange 
between non-relativistic electrons and ions
\cite{Braginskii:RPP1},
\begin{equation}\label{Peiclass}
P_{ei}^{(cl)}=-4\sqrt{2\pi}\,e^4 n_e m_{e0}^{1/2}
m_i^{-1}n_i Z_i^2\ln\Lambda\,\dfrac{T_e-T_i}{T_e^{3/2}}.
\end{equation}
From \eref{Pei3}, one can obtain the ultra-relativistic limit 
\cite{LifshitzPitaevskii_PhysKinetics1981}, 
$1\ll T_e/m_{e0}c^2 \ll\sqrt{137\ln\Lambda}$
(the upper boundary is defined by the validity of the small-angle 
scattering approximation \cite{BeliaevBudker:DAN1956mix}):
\begin{equation}\label{Peiultra}
P_{ei}^{(ur)}\simeq
-\dfrac{4\pi e^4 n_e n_i Z_i^2\ln\Lambda}{m_{e0}^{1/2}m_i c}\,
\dfrac{T_e-T_i}{T_e}.
\end{equation}
%

%%%%%%%%%%%%%%%%%%%%%%%%%%%%%%%%%% Fig01 start
\begin{figure}[!tb]
\begin{center}
\includegraphics[width=8.0truecm,angle=0]{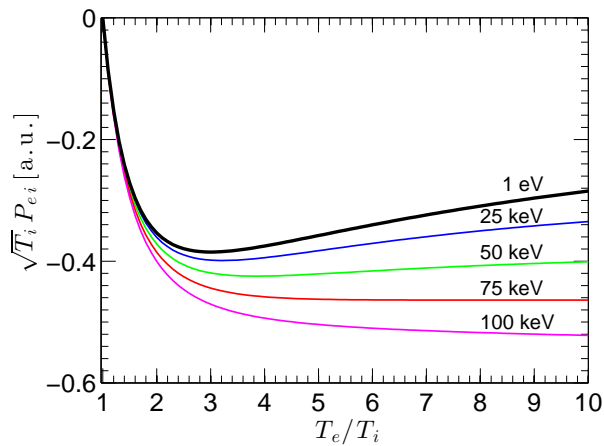}
\end{center}
\caption{(color online) The rate of energy exchange between the relativistic 
electrons and the ions, multiplied by $\sqrt{T_i}$, is shown as a function of 
$T_e/T_i$ for different $T_i$. 
The case $T_i=1$~eV is taken as the non-relativistic limit.}
\label{peimain}
\end{figure}
%%%%%%%%%%%%%%%%%%%%%%%%%%%%%%%%%%  Fig01 end

In \fref{peimain}, $\sqrt{T_i}P_{ei}$ is plotted as a function of 
$T_e/T_i$ with $T_e\geq T_i$ for different values of $T_i$ starting from 
the non-relativistic limit with 1~eV (shown with the bold line). 
For convenience of graphical representation of $P_{ei}$, the scaling 
factor $\sqrt{T_i}$ is applied, which makes $\sqrt{T_i}P^{(cl)}_{ei}$ 
a function only of the temperature ratio $T_e/T_i$.
One can see that for high temperatures the difference between $P_{ei}$ and 
$P_{ei}^{(cl)}$ becomes significant in the range $T_e/T_i > 2$, since the 
maximum of $|P_{ei}|$ shifts to the upper values (the shape of $P_{ei}^{(cl)}$ 
does not depend on the temperature).
For sufficiently high temperatures, $T_{e,i}>75$~keV, $P_{ei}$ becomes 
a monotonic function of $T_e/T_i$, and the extrema disappears.
The latter feature (absence of any extrema in $P_{ei}$ for high temperatures) 
is in agreement with the ultra-relativistic limit \eref{Peiultra}, 
$P_{ei}^{(ur)}\propto -(T_e-T_i)/T_e$, which is monotonic and has the same 
sign of the slope as $P_{ei}^{(cl)}$ in the range $T_e/T_i<3$.
This is a qualitative difference from the non-relativistic limit and 
the consequences are considered in the next section.

%=======================================================================
\section{Stability of the Coulomb coupling}
\label{sec:coupling}
%=======================================================================

For any set of plasma parameters which corresponds to the steady state, 
$P_{ei}+P_\mathrm{ext}-P_\mathrm{loss}=0$, the energy-balance equation 
\eref{e-balance} yields the following: 
if $P_{ei}^\prime\equiv dP_{ei}/dT_e>0$, 
this state is potentially unstable with respect to collisional decoupling, 
i.e. to growth of $T_e/T_i$ with decreasing $P_{ei}$.
Physically, this means that in plasmas where the ions are heated predominantly 
by the drag with hot electrons, positive feedback can appear 
\cite{BraunEmmert_NF1984,WagnerStroth_Transport-expview:PPCF1993,Zhang_HT7:JPP2010,
Maassberg_eroot:PoP2000,Rome_eroot:PPCF2006} 
which leads to a rapid transition to a new steady state with $T_e\gg T_i$ 
defined by other factors.
As a consequence, any further increasing of the electron heating power leads 
to further growth of $T_e$ (but not $T_i$) and might even provoke a thermal 
collapse of the ions.

From the condition $P_{ei}^\prime<0$, one can see that stability of 
the collisional coupling in non-relativistic plasmas is guaranteed by 
a very simple relation: $T_e<3T_i$.
For arbitrary temperatures, when relativistic effects are non-negligible, 
this condition is more complicated,
\begin{equation}\label{TeTi}
T_e<\left(3+\dfrac{2y}{1-y}\right)T_i,
\end{equation}
where
\begin{equation}\label{y}
y(\mu_r)=-\mu_r\left(\dfrac{K_3(\mu_r)}{K_2(\mu_r)}-1\right)+\dfrac{5}{2}+
\dfrac{2(\mu_r+2)}{\mu_r^2+2\mu_r+2}.
\end{equation}
Since $y\simeq (8\mu_r)^{-1}$ for $\mu_r\gg 1$ and $y(c\rightarrow\infty)=0$, 
\eref{TeTi}, as expected, recovers the non-relativistic limit.

\begin{figure}[!tb]
%\vskip -0.5cm
%\hskip -0.75cm
\begin{center}
\includegraphics[width=9.0truecm,angle=0]{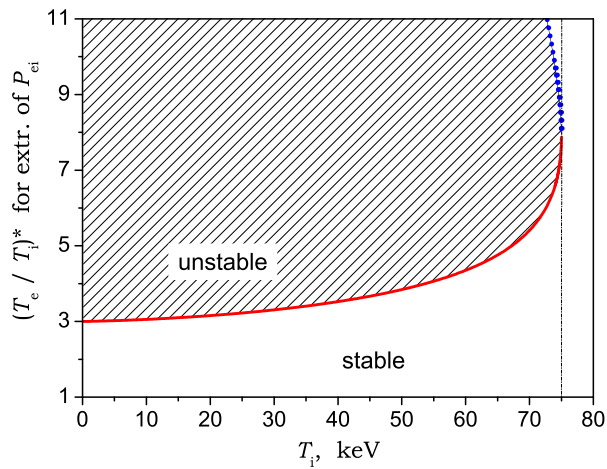}
\end{center}
\caption{(color online) Ratio $(T_e/T_i)^\star$ which corresponds to 
the maximum (full (red) line) and minimum of $|P_{ei}|$ (dotted (blue) line). 
The area where collisional decoupling is impossible is marked as 
``stable'' and vice versa.
}
\label{peimax}
\end{figure}

The analytical investigation of \eref{TeTi} is rather cumbersome, but can 
easily be done numerically. 
In \fref{peimax}, the solution of \eref{TeTi} is represented as the ratio 
$(T_e/T_i)^\star$ which corresponds to the extrema of $P_{ei}$ plotted 
as a function of the ion temperature $T_i$. 
One can see that, starting from the low temperature limit $(T_e/T_i)^\star=3$, 
the first extremum (maximum of $|P_{ei}|$) shifts towards higher values with 
increasing $T_i$ and the extrema disappear for $T_i>75$~keV, where 
$P_{ei}(xT_i;T_i)$ becomes a monotonic function of $x=T_e/T_i$. 
In the area where $P_{ei}^\prime>0$ (labelled ``unstable'' in \fref{peimax}), 
the collisional decoupling can easily appear since any increase of the 
electron temperature leads to a degradation of the collisional energy transfer 
to ions, creating positive feedback. 
If other factors do not prevent this feedback, a new steady state with 
$T_e\gg T_i$ must be established.
In the opposite case, i.e. if $P_{ei}^\prime<0$, the present steady state is 
absolutely stable and no decoupling can arise. 
In \fref{peimax} this area is labelled ``stable''.
%
%%%%%%%%%%%%%%%%%%%%%%%%%%%%%%%%%% Fig02 start
%%%%%%%%%%%%%%%%%%%%%%%%%%%%%%%%%%  Fig02 end
%

%=======================================================================
\section{Summary and discussion}
\label{sec:summary}
%=======================================================================

In this paper, the condition for steady-state stability with respect to 
the collisional decoupling between electrons and ions in hot plasmas for 
a broad range of temperatures has been defined in the relativistic approach.
It was shown that while the stability condition in non-relativistic plasmas 
is given by $T_e/T_i<3$, relativistic effects make this threshold dependent 
on the temperature and shift its value to higher $T_e/T_i$ with an increase 
of temperature.
For temperatures $T_{e,i}>75$~keV, the collisional coupling becomes 
absolutely stable for any temperature ratio. 
This result can be useful for interpretation of experiments with heating 
of plasmas in the electron channel, when the ion heating is caused 
exclusively by the collisional energy exchange with electrons.

Collisional decoupling appears only if the heating of electrons is 
sufficiently high and the temperature threshold can be reached.
Physically, this means that the heating of electrons has a power threshold 
with respect to decoupling, but its value is defined by the concrete
scenario and is not discussed here.

Since the consideration here is restricted to the collisional energy 
exchange between electrons and ions, the energy loss of electrons due 
to the radiation (in particular, the bremsstrahlung, which accompanies 
collisions), was excluded from consideration. 
In fully ionized low temperature plasmas, the contribution of radiation 
in the energy balance is negligible, but for higher temperatures this 
phenomena must be taken into account.
Moreover, if conditions for the collisional decoupling appear for 
sufficiently high temperatures, the losses through radiation, 
$P_{loss}=P_{Bs}+P_{ce}$ (here, $P_{Bs}\propto \sqrt{T_e}$ is bremsstrahlung 
and $P_{ce}\propto T_e$ is cyclotron radiation), 
produce a stabilizing effect for electrons by counteracting the increase 
of $T_e$. 
However, the radiation does not have any direct influence on the power 
balance for ions.
Thus, the radiation does not change the temperature threshold for 
the collisional decoupling even for high temperatures, but can be important 
for the power threshold.

%=======================================================================

\begin{ack} 
The authors would like to acknowledge Per Helander for support and 
fruitful discussions. 
\end{ack} 

%=======================================================================
%\vskip 12pt
\section*{References}
\bibliographystyle{prsty}
\bibliography{pei_biblio}

%==========================================================================
\end{document}